\documentclass[preprintnumbers,prd, 10pt, amsmath,amssymb,a4page,twocolumn, latexsym,array,enumerate,letter,numbers]{revtex4}
\usepackage{graphicx}% Include figure files
\usepackage{dcolumn}% Align table columns on decimal point
\usepackage{bm}% bold math
\usepackage{amssymb}
\usepackage{epsfig}
\usepackage{slashed}
\usepackage{color}

\usepackage[
       colorlinks=true,
      filecolor=black,
       anchorcolor=blue,
      linkcolor=blue,
      citecolor=cyan,
      urlcolor=cyan,
       linktocpage=true,
        plainpages=false,
        breaklinks=true,
            pdfstartview=FitH
           ]{hyperref}

\def\cm{\color[rgb]{0.00,0.6,0.00}}
\def\cm{\color[rgb]{0.00,0.0,0.00}}

\def\Rs{R_s}

\def\XT{ X{\scriptsize ENON}1T}

\begin{document}
\title{Dark fluxes from Accreting Black Holes through Several Mechanisms} 
%\title{Dark fluxes from Accreting Black Holes and  Direct Detections} 
%Matter and Xenon1T
%\title{Axion Flux and Fast Dark Matter from Accreting Black Holes}
%Black Hole Accretion }  %and axion Fast No. 19A Yuquan Road,
%\title{ Fast Dark matter and Axion from Supermassive Black Holes and Xenon} Supermassive
%\title{Black hole sources for keV dark matter, dark photon and axion fluxes } Active Galactic Nucleus and
\author{Rong-Gen Cai$^{1,2,3,9}$}%\email[]{cairg@itp.ac.cn}
\author{Sichun Sun$^{4,5}$} %\email[]{Sun.Sichun@roma1.infn.it}%corresponding author
\author{Bing Zhang$^{6,7}$}%\email[]{zhang@physics.unlv.edu}
\author{Yun-Long Zhang$^{8,2,9,10}$}%\email[]{yun-long.zhang@yukawa.kyoto-u.ac.jp}%corresponding author

\affiliation{$^1$CAS Key Laboratory of Theoretical Physics, Institute of Theoretical Physics, Chinese Academy of Sciences, Beijing 100190, China}
\affiliation{$^2$School of Fundamental Physics and Mathematical Sciences, Hangzhou Institute for Advanced Study,
University of Chinese Academy of Sciences, Hangzhou 310024, China}
\affiliation{$^3$School of Physical Sciences, University of Chinese Academy of Sciences, Beijing 100049, China}
\affiliation{$^4$School of Physics, Beijing Institute of Technology, Haidian District, Beijing 100081, China}
\affiliation{$^5$Department of Physics and INFN, Sapienza University of Rome, Rome I-00185, Italy}
\affiliation{$^6$Nevada Center for Astrophysics, University of Nevada, Las Vegas, NV 89154, USA}
\affiliation{$^7$Department of Physics and Astronomy, University of Nevada, Las Vegas, NV 89154, USA}
\affiliation{$^8$National Astronomy Observatories, Chinese Academy of Science, Beijing, 100101, China}
\affiliation{$^9$International Center for Theoretical Physics Asia-Pacific, Beijing/Hangzhou, China}
\affiliation{$^{10}$Center for Gravitational Physics, Yukawa Institute for Theoretical Physics,  Kyoto University, Sakyo-ku, Kyoto 606-8502, Japan}

\begin{abstract}\cm
We discuss the possibility that accreting black hole systems can be sources for dark matter flux through several different mechanisms. We firstly discuss two types of systems: coronal thermal plasmas around supermassive black holes in active galactic nuclei (AGNs), and accretion disks of stellar-mass X-ray black hole binaries (BHBs). We explore how these black hole systems may produce keV light dark matter fluxes and find that the dark fluxes from those sources might be too weak to account for the current XENON1T excess.
On the other hand, black holes can be good accelerators to accrete and boost heavy dark matter particles. If considering collisions or dark electromagnetism, those particles can then escape and reach the benchmark speed of 0.1c at the detector. We also extend the black hole mass region to primordial black holes (PBHs) and discuss the possibility of contributing to keV light dark flux via superradiance of PBHs.
\end{abstract}
%or Hawking radiation 
%\preprint{YITP-20-114}
\maketitle
 \tableofcontents

\section{Introduction}

Dark matter direct search experiments have been very successfully developed to put constraints on the dark matter properties \cite{XENON:2020rca}.  While the search for a few GeV dark matter is still going on without confirmed signals so far, lighter sub-GeV dark matter scenarios have received more attention in recent years \cite{Boehm:2003ha,Feng:2008ya,Knapen:2017xzo,Essig:2011nj,Knapen:2016cue,Hochberg:2016ntt,Capparelli:2014lua,Acanfora:2019con,Caputo:2019cyg,Caputo:2019xum}, encouraged by the comic ray excesses \cite{Jean:2003ci, Jeltema:2014qfa, Adriani:2008zr}. The galactic center dark matter has been widely considered to be the sources of the cosmic ray excesses as well as the possible explanation for the {\XT} excess \cite{XENON:2020rca, Zhang:2020htl, PandaX-II:2020udv}. The other sources for relativistic dark matter particles or axions are the Sun and stars, although such light dark matter models face many astrophysical constraints such as stellar cooling \cite{DiLuzio:2020jjp}.

% there are some matters Outside the event horizon and outside the Innermost Stable Circular Orbit (ISCO), the black hole hasThe ISCO is the inner circle of the accretion disk, $R_{ {I}}=2\Rs$.
%Most of the compoents are in X-ray radiations, around 1-100 keV.
%. This is very close to the absolute zero and

With the first detection of gravitational waves from the binary black hole mergers \cite{Abbott:2016blz} and the Event Horizon Telescope (EHT) \cite{Akiyama:2019cqa}, we have new methods to probe black holes. A black hole can be a host of many processes involving the beyond the Standard Model (SM) physics \cite{Rosa:2017ury,Sun:2020gem}.  Particles may be accreted and pairs may be produced around the event horizon, which can then be accelerated around the black hole without coupling to SM physics. In principle, the dark particles can also be thermally produced or thermalize around the black holes, and be ejected along with the standard keV X-ray radiations. {\cm  The field of dark matter has very rich and diverse contents, here we just choose three difference types: light ($m \sim$ eV) dark photon/axion-like particles, intermediate ($m\sim$ keV) dark photon dark matter~\cite{Nelson:2011sf,An:2020bxd} and heavy ($m\sim$ MeV to GeV) heavier dark matter. We may further specify the details of these viable models, e.g. with/without self-interactions, and with/without the couplings to the SM models.  Since all these models can be quite different, and there are so many different workable types in the literatures (there are even multiple models to explain Xenon1T), we try our best to organize these possibilities and be inclusive. }
 
%For the dark matter, no matter the couplings to the standard model, they will get pair produced around the event horizon. % \cite{?}. black hole orbit
One can see related examples like how to produce dark matter relic in the early universe through black hole evaporation in \cite{Lennon:2017tqq, Hooper:2020evu,Baldes:2020nuv}. Black holes have also been discussed as the arbitrary high energy particle accelerators \cite{Banados:2009pr,Jacobson:2009zg}. Moreover, the disks around accreting black holes can be the places for nucleosynthesis \cite{hu2008nucleosynthesis}. The visible  accretion disk is heated through viscous dissipation of gravitational energy. 

\begin{figure}[h]
\centering
%\raggedright
\includegraphics[scale=0.23]{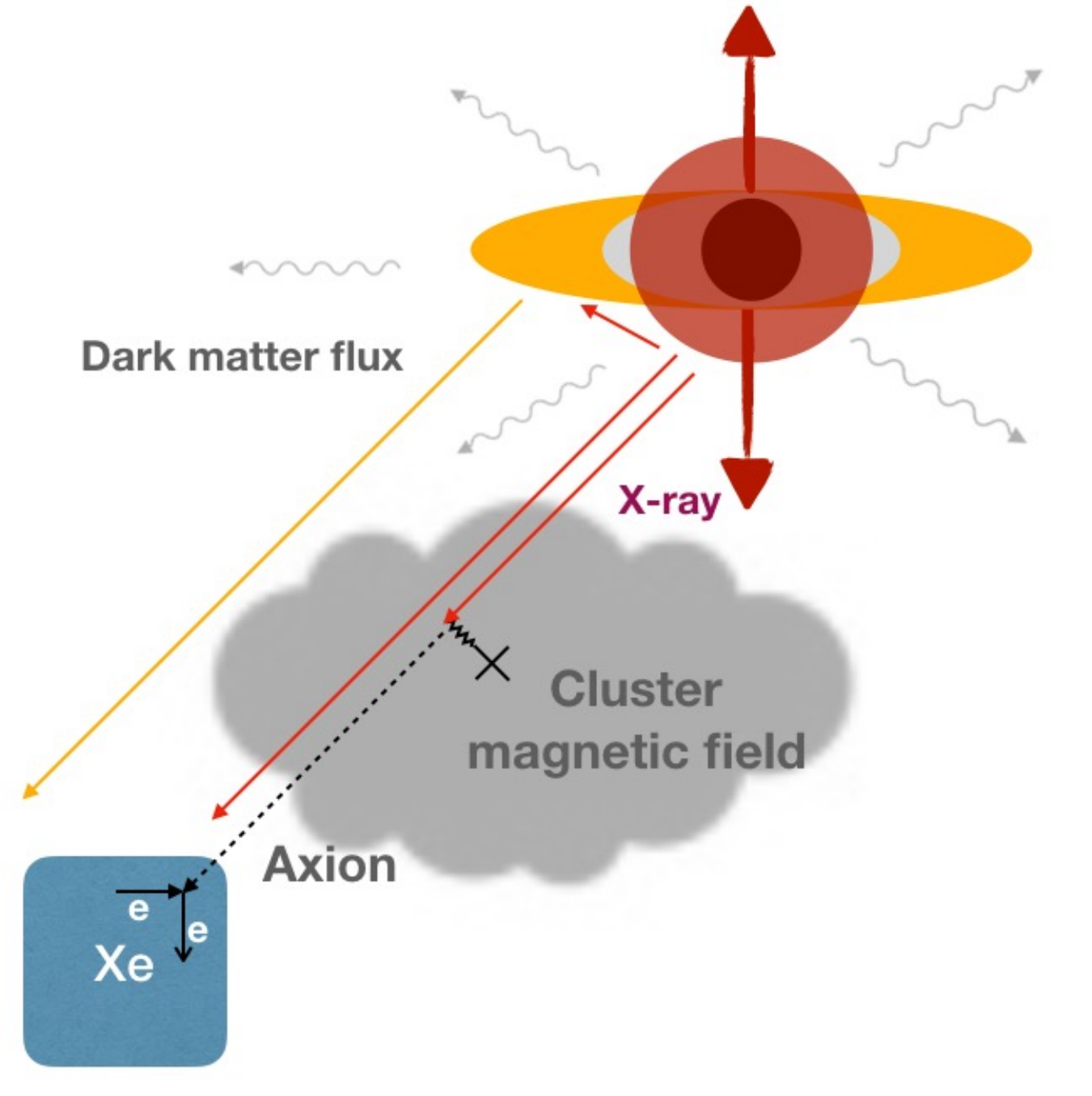}  \\
\caption{The schematic diagram of axion flux converted from the X-rays passing through the cluster of magnetic field.\label{figANG} }
\end{figure}

For light particles, such as dark photon or axion which may be thermally produced in the coronal thermal plasmas and accretion disks,
a flux of them at keV temperature would be expected, similar to the solar source \cite{Redondo:2013wwa}.  Astrophysical black holes are known to carry a large-scale magnetic field, which is the agent to tap the spin energy of black holes through the Blandford-Znajek process. The poloidal magnetic field of a black hole provides an environment for photon-to-axion conversion ($\gamma \rightarrow a$) \cite{Blandford:1977ds,Ackerman:2008kmp}. Besides, as depicted in Figure \ref{figANG}, the cosmic magnetic field between the source and Earth can convert part of the X-ray photons to axions at the same energy scale, which is similar to neutrino oscillations.  It has been found that up to 1/3 of photons can be converted, when the traveling distance inside the magnetic field is saturated \cite{Burrage:2009mj,Pettinari:2010ay,Mirizzi:2006zy}. 
 
%Active Galactic Nuclei (AGN) with coronal plasmas  (red region $\sim $keV) and the cold accretion disk (orange region $\sim 10^{-2}$ keV)

If couplings between the dark sector and SM fields are quite small (For e.g. emergent dark sector in \cite{Cai:2017asf}, there is only gravitational interaction between the dark sector and the standard model sector.), the thermal production of dark particle radiation from the visible thermal media, if any,  is much smaller than the radiations of the Standard Model charged particles.  %For the case of light dark matter without the dark electromagnetism (which is presumably mediated by the light dark photons), 
 For this case, we can look back at the traditional Penrose process \cite{Penrose:1971uk} or superradiance, where energy can be extracted from the rotating black holes directly.  Axions/dark photon vectors may also be accumulated around the black hole through gravitational superradiance effects and form clouds \cite{Arvanitaki:2009fg}. 
The process can also be independent of the black hole accretion of standard model matters.

The heavier MeV - GeV dark matter particles can also be accelerated to be relativistic, since the orbiting speed at the so-called Innermost Stable Circular Orbit (ISCO) is around the order of 0.1c.  These kinds of sources with keV kinetic energy are a lot less constrained compared to stellar sources, and provide a mechanism to accumulate large dark matter density.
%much larger than the Galactic center \cite{Buch:2020mrg}.
% such that the electromagnetism related energy extraction which includes the process that the charged particles are
%If one assumes the existence of the dark electromagnetism, the dark Blandford-Znajek process may be relevant.
Moreover, black hole systems provide a good mechanism to heat up dark matter in this heavier dark matter case. The scattering process around the black hole can also heat up dark particles around it, in analogy to the visible process considering that black holes can be assumed to interact with the visible/dark sector equivalently. In this case, we need self-interactions in the dark sector or couplings to the standard model to provide enough scattering. In this self interacting dark matter case \cite{Tulin:2017ara}, accretion disk model with the viscous hydrodynamical approximation can be applied,  e.g. the double-disk dark matter \cite{Fan:2013yva,Fan:2013tia}. The cross section of self interacting dark matter is bounded by $  \sigma _{\chi}/m_{\chi} \lesssim 1 \text{cm}^2/\text{g}$ \cite{Randall:2007ph}. It is however, comparable to $  {\sigma_T}/{m_p} \simeq 0.3  \text{cm}^2/\text{g}$ for ionized hydrogens with the Thomson scattering cross-section. We will show how the dark luminosity can be as bright as the visible luminosity in a model made of dark photon and dark fermion later.

%More evidences support the viscosity of dark matter models in recent years.

In the following sections, we discuss these processes in more details and relate them to the{\XT} excess explanations. In section \ref{Xray}, we discuss the case with the light ($m \sim$ eV) dark matters, including axions and dark photons, which are converted from the X-rays passing through the cluster of magnetic field.
In section \ref{Flux}, we discuss other possibilities of the dark processes around black holes, 
%In section \ref{others}, we discuss 
including self-interaction dark matter from the accreting black holes,
heavier dark matter boosted from accreting black holes, 
and superradiance from primordial black holes (PBHs). 
We summarize and discuss these issues more in section 
\ref{Con}.
 %can also produce dark particles/axions. However, in order to explain the Xenon 1T excess in the low energy electron recoil between 2 - 4 keV, the fluxes from the known source of AGNs are too weak, but there are still possibilities for the BHBs or the primordial black holes (PBHs) inside the Milky Way galaxy. 

\section{Dark Matter Produced from keV Environment near Black Holes}
%\section{KeV  Dark Matter and X-ray Black Holes} %emissions and  Light
\label{Xray}

In this section, 
%we discuss different kinds of host black holes with the temperature of keV environment.
we discuss several environments associated to accreting black holes of different masses and characterized by keV-scale temperatures.
%One class is the active galactic nuclei (AGNs) of supermassive black holes which are heavier than 
We first consider Active Galactic Nuclei, characterized by supermassive black holes heavier than
$10^5M_{\odot}$.  The other class is the stellar-mass black holes ($3-10^2M_{\odot}$) with companion stars, which are the so-called X-ray black hole binaries (BHBs). There are also possibilities for primordial black holes (PBHs) such that the mass range can be extended to the region that is much smaller than three solar masses.

\subsection{KeV environment around the black holes} 
The observations and models show that the radiations from plasmas around black holes are partly X-rays, around the keV scale. The light dark matter particles/axions can be thermally produced around visible matter and their energy is expected to be around keV,  in analogy to the production in the Sun. The keV medium can produce axion/dark photon flux, as calculated in \cite{An:2013yua, Redondo:2013wwa}, through a few processes.
These processes include dark Higgs strahlung, oscillations from visible photons to dark photons, Primakoff processes $\gamma+Ze\to Ze+a$, and ABC processes for axion production. The ABC processes stand for Atomic axio-recombination and Atomic axio-deexcitation, axio-Bremsstrahlung in electron-Ion or electron-electron collisions, Compton scattering (see eg \cite{An:2013yua} for a sample of Feynman diagrams).

%Orbits Around the Black Holes}
% \subsection{Photon sphere}% are traveling in speed of light in the orbits.Accreting

The Eddington luminosity corresponds to the balance between gravity and radiation in the spherical plasmas  i.e. \cite{Astrophysics1979}. It leads to $\frac{G M m}{R^2}=\frac{L_E}{c}\frac{\kappa_E m}{4\pi R^2}$, where $m$ is the test mass of a small part in the accreting matter at radius $R$. One then obtains the Eddington luminosity
\begin{align}\label{EL}
L_E= {4\pi G M c } /\kappa_E,
\end{align}
where $\kappa_E\equiv {\sigma_i}/{m_i}$ is the cross section due to photon scattering per unit mass. In the high energy approximation of accretion, the accreting matters are mostly composed on ionized hydrogen  and the mass is dominated by the proton mass $m_p$.
The opacity is provided by  Thomson scattering with the cross-section
% is dominated by the radiation pressure on the electrons 
$\sigma _{\rm {T}}$. %with the Thomson scattering cross-section .

Observationally  \cite{Arevalo:2014zva,Parker:2014kna}, astrophysical black holes radiate keV X-rays, which can be explained in theoretical models.
If most of the emission is released in the keV band, one may estimate the X-ray flux from the Eddington luminosity in \eqref{EL}, which leads to the total energy flux % ${\mathcal{E}_E}$,
\begin{align}\label{PE0}\!\!
{\mathcal{E}_E}&\equiv \frac{L_E}{4\pi R^2 } \simeq  \left( \frac{M}{M_{\odot}}\right)\left(\frac{\text{kpc}}{R}\right)^2\!\!\times 10^3 \text{keV} /(\text{cm}^{2}\cdot\text{s} ). %  \times
\end{align}
The number flux can be obtained through ${\Phi_E} \equiv {\mathcal{E}_E}/E$,
where $E$ is the energy scale of each particle in the flux.  For super massive black hole at kpc distances, the energy flux ${\mathcal{E}_E}$ is comparable to the X-ray flux from the Sun detected at the Earth at $1 \text{ erg}/(\text{cm}^{2}\cdot\text{s}) \simeq 10^{8}  \text{keV}/(\text{cm}^{2}\cdot\text{s})$.  However, most of those active AGNs are at Gpc distances. Since their luminosities are much higher, their fluxes can be comparable to those of X-ray BHBs as discussed below.

\vspace{5pt}
{\it 1. AGNs in supermassive black holes.}
One source of X-ray radiations is the active galactic nucleus (AGN), which is a compact region at the center of a galaxy that has extremely high luminosity. %The wide waveband
Broad-band emission from radio wave of $O(10^{-3})$ keV to gamma-ray of $O(10^{3})$ keV has been observed, which is powered by accretion of matters surrounding supermassive black holes ranging from $10^5M_{\odot}$ to $10^9M_{\odot}$. For supermassive black holes with typical mass $M\sim 10^6 M_{\odot}$, the luminosity is $L \sim 10^{44}$ erg/s.  At the galactic center of the Milky Way galaxy, there is a supermassive black hole with mass $4\times10^6M_{\odot}$ which powers the compact radio source Sgr A* and is about  $8$ kpc from Earth \cite{Gillessen:2008qv,Gillessen:2013}, which might be a possible source for the dark matter fluxes. 
In our Galactic center, the luminosity of Sgr A$*$ \cite{Corrales:2020jin, Manshanden:2018tze} and other low luminosity AGNs is in the order of $10^{-3}L_E$. For very active galaxies, the bolometric luminosities can be higher than $10^{2} L_E$.

\vspace{5pt}
{\it 2. X-Ray Black Hole Binaries (BHBs).}

Except for the matter accretion of AGNs, around stellar-mass black holes with stellar companions, the accretion of matter is also happening. They are so-called X-Ray black hole binaries (BHBs).
%Accreting black holes in the Galactic scale are BHBs.
%Hailey, C., Mori, K., Bauer, F. et al. 
The candidates include Cygnus X-1, XTE J1650-500 and GX 339-4 \cite{Hanke:2008ia,Miller:2003wr,Gallo:2008iz}.
A density cusp of quiescent X-ray binaries can also be found in the central parsec of the Galaxy \cite{Galaxy2018}. 
There are two kinds of  X-ray spectra of the accreting black holes. In the so-called high-soft state, the X-ray spectrum is composed of a strong and narrow peak at a few keV and a soft power-law spectrum up to hundreds of keV. The low-hard state peaks below 1 keV while the Comptonized component is extended up to 100 keV. The luminosity of the high-soft state can be near or much higher than $L_E$, while the low-hard state is usually  two orders of magnitude below the Eddington luminosity $L_E$.

The widely considered accretion flow model is the so-called Shakura-Sunyaev model \cite{Shakura:1972te}, which is used for the observed emission energies less than 10 keV.
This thin disk model was also worked out by Lynden-Bell, Pringle and Rees \cite{LyndenBell:1974kk},
and a general relativistic treatment for the inner part of the disk can be found in Novikov and Thorne's  \cite{Novikov:1973, Page:1974he}. For the purpose of order of magnitude estimation of this paper,
% the simple Shakura-Sunyaev disk model suffices.  
we take the Shakura-Sunyaev model as an example to show that they can produce keV emission. There are many more variants of accretion models. As long as they can produce X-rays, our following discussion remains relevant.
We use the notations $r=R/R_s$ and $m_{BH}=M/M_{\odot}$. The local radiation energy flux from unit surface at the radius $R$ of the disk is determined by the gravitational energy release ${\mathcal{E}_r} =\frac{\mathcal{E}_0}{r^3}\left( 1- \sqrt{ 3/r } \right)$,
where $ {\mathcal{E}_0} = \frac{3}{16\pi} \frac{\dot{M} c^2}{\Rs^2} $.
%\begin{align}\label{FR} {\mathcal{E}_r} =\frac{\mathcal{E}_0}{r^3}\left( 1- \sqrt{ 3/r } \right),\quad {\mathcal{E}_0} = \frac{3}{16\pi} \frac{\dot{M} c^2}{\Rs^2} \,.\end{align}
The luminosity is related to the accreting rate via
$L={f_s} \dot{M}c^2$, where ${f_s}\simeq0.06$ for the Schwarzschild black hole and ${f_s}\simeq0.4$ for the extremal Kerr black hole \cite{Pringle:1981ds}.

\begin{figure}[h]
\centering
%\raggedright
\includegraphics[scale=0.45]{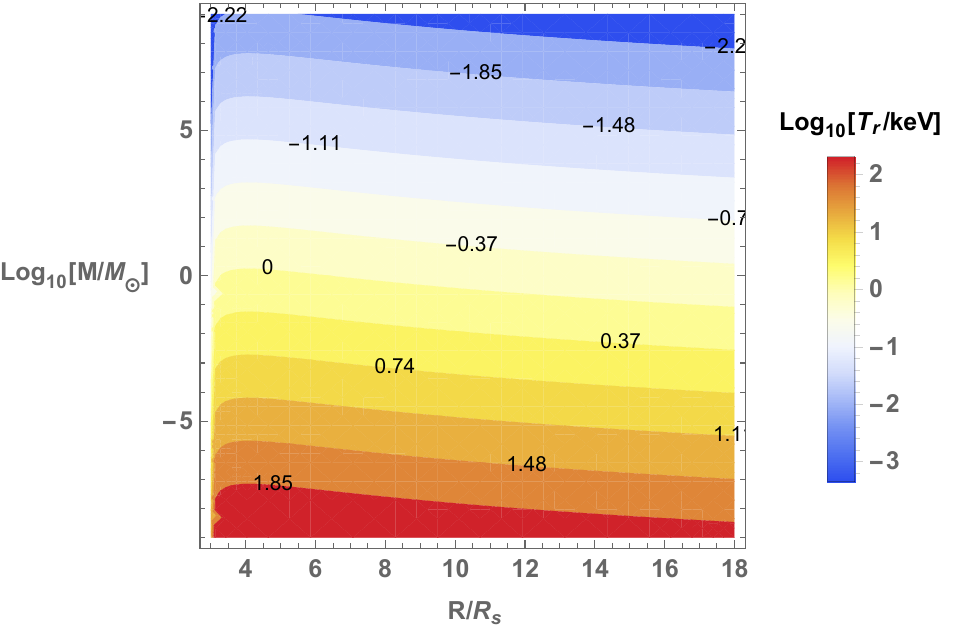}\qquad \\
\caption{The contour plot of temperature ${T_r}$ in \eqref{TR} of thin accretion disk as a function of locations in the disk $R/R_s$ and black hole masses $M/M_{\odot}$, where $f_e=1$ and $f_s=0.1$ have been taken.} \label{figT}
\end{figure}

If considering the critical Eddington luminosity with $L={f_s} \dot{M}c^2=f_e L_E$, the typical energy flux is around ${\mathcal{E}_0}  = \frac{3 f_e }{16\pi f_s } \frac{L_E}{\Rs^2} $. %where  $f_e=1$ and $f_s=0.1$ will be taken,
Here the Eddington luminosity has been approximated as $ L_{E} \simeq 3.2 \times 10^4 \left(\frac{M}{M_{\odot}}\right) L_{\odot},$
where  $L_{\odot}\simeq 3.8\times10^{33}$ erg/s is the solar luminosity.
%\begin{align}\label{EL}
% L_{E} %&\simeq  1.26 \times  10^{38}\left(\frac{M}{M_{\odot}}\right) \text{erg/s}\nn\\
% &\simeq 3.2 \times 10^4 \left(\frac{M}{M_{\odot}}\right) L_{\odot},
%\end{align}
In the following calculation, we will simply take  ${f_s}\simeq0.1$ and  $f_e=1$ for the estimation \cite{Bambi:2018thh}.
The Shakura-Sunyaev disk is optically thick, which can be roughly approximated as blackbody.
From the law of thermal radiation $\sigma_s T_{r}^4={\mathcal{E}_r}$ with the Stefan-Boltzmann's constant  $\sigma_s=\frac{\pi^2k_B^4}{60\hbar^3 c^2}$, the local effective temperature of the disk becomes
\begin{align}\label{TR}
T_{r} = \frac{( 1- \sqrt{3/r} )^{1/4}}{r^{3/4}}T_{0}, \quad
T_{0}  \simeq   \left(\frac{10}{m_{BH}}\right)^{1/4} \!\!\!\times {  3\,\text{keV} }.
\end{align}
The typical value of $T_{0}$ is calculate from $T_0\equiv\left( \mathcal{E}_0/{ \sigma_s  }\right)^{1/4}$.
% \simeq   {  3\,\text{keV} }\big[ {f_e}/{(mf_s) }\big]^{1/4}

In Figure \ref{figT}, we make the contour plot of temperature ${T_r}$ in \eqref{TR} as a function of locations in the disk $r=R/R_s$ and black hole masses $m_{BH}=M/M_{\odot}$. One can see that the peak temperature of the disk is around ${T_r} \, {\simeq}\,1.2\,\text{keV} /m^{1/4}$ at the radius $r\simeq 4$.
And in Table \ref{tableBH1}, we list two benchmark points for black hole sources.
The gravity of primordial black holes is too weak to have significant accretion, so we leave the discussion to section \ref{others}.
%\ \sim 10^{33} /\left( {f_s} m\, r^3 \right) \text{keV}\cdot {\text cm}^{-2}\cdot \text{s}^{-1}
%In X-ray binaries it peaks at energies of 1 keV, while for AGNs around 10 eV.
% and the factor $1/2$ is considerig one side of the disc only \cite{}.
%=  \left( {J_R} /\sigma_s\right)^{1/4}
%hen $T_{r}\propto 1/M^{1/4}$ after considering \eqref{FR} and \eqref{TR}, or

 \begin{table}[th]
\begin{tabular}{ |c|c|c| }
\hline%/$(\text{cm}^{2}\cdot\text{s}\cdot \text{keV})$
Types of X-ray sources & AGNs & BHBs    \\
\hline
Benchmark Masses ($M_{\odot}$) & ${\sim}10^6$ ~ & ${\sim}10~~$   \\ %\gtrsim
\hline
Eddington Luminosity $L_E$(erg/s) & ${\sim}10^{44}$  & ${\sim}10^{39}$   \\
\hline
Disk Temperature $T_0$(keV)  & ${\sim}10^{-2}$   &  ${\sim}1~~$   \\
%\hline Distances(kpc) &$ <1$ & $<10^{-3}$  & $<10^{-6}$ \\
\hline
\end{tabular}
\caption{ Here we list two benchmark points for accreting black hole sources, with the Eddington luminosity and the peaked temperature in thin accreting disk in \eqref{TR}.
%and require ${\mathcal{E}_E}>10^{10} \text{keV}/(\text{cm}^{2}\cdot\text{s})$ for possible Xenon1T explanations of the axion like particles or dark photons in Section \ref{Flux}.
\label{tableBH1}}
\end{table}
 %with a

With the simple model of the thin accretion disk with the temperature distribution in \eqref{TR}, in the X-ray black hole binaries, the temperature of the disk peaks at energies around 1 keV. The temperature of disk around the supermassive black hole peaks at around $10^{-2}$ keV.  For AGNs, more complicated models are required. From the observations of a couple of AGNs, e.g. 1H 0707-495 \cite{Zoghbi:2009wd,Wilkins:2012zm,Wilkins:2011kt} and analysis, the X-ray emission comes from the corona surrounding the black hole of an AGN as well as the reflection and thermal emission of the rays from the accretion disk. It has multiple features including reflection from accretion disk from 0.5 keV to 1 keV,  corona X-ray around 1-5 keV and Iron K$\alpha$ emission line peaked at $\simeq 6.8$ keV. The number flux in the spectrum from sources~\cite{Zoghbi:2009wd,Wilkins:2012zm,Wilkins:2011kt} is around $\frac{d\Phi}{dE}  \simeq 10^{-3}  \text{photons  }/(\text{cm}^{2}\cdot\text{s}\cdot \text{keV} )$.

\subsection{Generating keV dark matter}

The temperature of AGNs or %disk around the X-ray binaries
X-ray binary disks can be as high as keV, which is close to the temperature in the core of the Sun. However, the density of the accretion disk for AGNs with mostly ions and electrons, $\rho_{\text{disk}} \sim10^ {-7} \text{g}/\text{cm}^3$
%which still depends on the different models
\cite{Garcia:2013oma},  is much smaller than the $\rho_{\text{HB}}\sim 10^4 -10^6\, \text{g}/\text{cm}^3$ (HB stands for the stars in the horizontal branch with the solar masses),  and is model-dependent.
%{\cm The BHB disk have a higher enough density for nuclear fusion to happen.}
The nuclear reactions take place in the core of the Sun and the reaction rates are highly temperature dependent. Black hole accretion disks can have high temperatures generated by the high accretion rates,
% with nuclear fusions, 
such that dark particles/axions can be generated in this environment.

%{\cm However,(Bing)
%in the core of Sun, the matter density is very high, which may allow such an equilibrium to be reached (photons have very high optical depth, so that the surface temperature of the Sun is much lower than the core temperature). The BHB disk may (or may not) have a high enough density to reach such a putative equilibrium state, but the AGN corona is likely too optically thin to make this happen. Otherwise, the solar corona (instead of the solar core) would be a much more powerful place to produce dark fluxes. }

%For the AGN corona, the density is rather low and the dark particles/axion generation is not that efficient and the dark luminosity is estimated to be a few orders of magnitude lower than the visible luminosity \cite{}.  %So the key is the produced dark photons/axions are of the same local temperature of the medium, which is a thermal effect that does not necessarily require thermal equilibrium.
Similar to the solar environment where the axion is converted via $\gamma \to a$  \cite{DiLuzio:2020jjp},  AGN corona (or solar corona) and X-ray BHB accretion disks can have a temperature of $\sim 10^7$ Kelvin and produce keV dark matter. Thermal emission is a process where the collected excitations, plasmons are converted to the light-dark matter of the comparable energy. The frequency of the plasmons is dominated by the medium temperature at a high temperature.  However, since the production rate is related to density so that the production rate from AGNs has been found to be quite low \cite{Jain:2009hf}. Only the solar corona or BHB accretion disks may reach an considerable axion flux, which is usually bound by the visible luminosity. Considering that the luminosity of dark particles/axions from the Sun is bound by $10\%$ of visible luminosity \cite{An:2013yua, Redondo:2013wwa}, we assume that the dark luminosity $L_D$ is comparable to the visible luminosity.
%However, AGN corona is likely too optically thin to make this happen,

For AGN corona, one needs to consider the alternative mechanism as shown in Figure \ref{figANG}.  It has been proposed that the cosmic magnetic field may convert part of the X-ray photons to high energy axions \cite{Burrage:2009mj,Pettinari:2010ay}, where
up to about 1/3 of photons can be converted, similar to neutrino oscillations. Thus, we simply start from the total energy flux \eqref{PE0} with Eddington luminosity in \eqref{EL}, and make the major assumption that the total dark luminosity is characterized by the Eddington luminosity  ${L_D}\lesssim {L_E}$.

% dark flux is bounded by ${\mathcal{E}_D}\lesssim {\mathcal{E}_E}$.

Notice here that this assumption ${L_D}\sim {L_E}$ is rather ad hoc. If the dark particles are produced thermally, the couplings with the
standard model particles are bound by the stellar cooling constraints which further constrains the production rate. %so it is reasonable to assume the similar bounds for the black hole productions with bounded couplings. 
If the dark flux is not produced thermally, then Eddington luminosity can be a benchmark point for the dark luminosity, assuming the dark sector is accreted by the black holes and balanced by some self-interaction-induced scatterings within the dark sector. If considering the axion fluxes converted from the X-ray photons under the cosmic magnetic field, this bound ${L_D}\sim {L_E}$ is also plausible.

\begin{figure}[h]
\centering
%\raggedright
\includegraphics[scale=0.45]{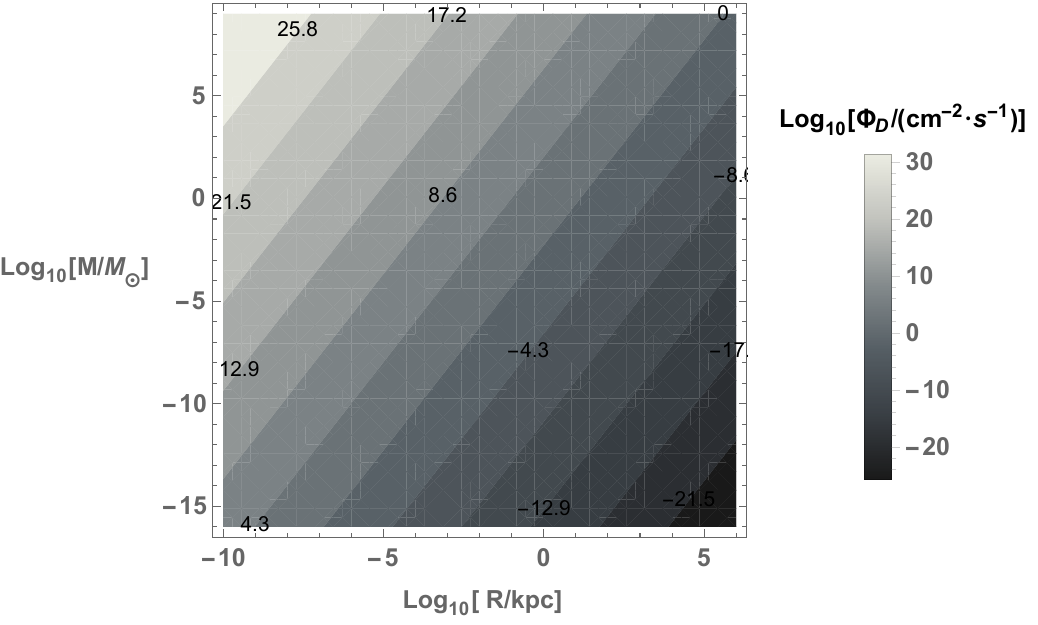}\qquad \\
\caption{The contour plot of the bound of total dark flux ${\Phi_D}$ in equation \eqref{PE} with the Eddington luminosity and $E_D\simeq 3.5$keV, as a function of the distances $R/$kpc and masses $M/M_{\odot}$ of black holes.\label{figFlux}}
\end{figure}

%\vspace{10pt}
In summary, the total number flux is given by ${\Phi_D}\equiv {\mathcal{E}_D}/E_D= {L_D}/{(4\pi R^2 E_D)} $ where $E_D$ the energy of dark matter particles. Considering the bound for the dark luminosity ${L_D}\lesssim {L_E}$, the number flux is bound by
\begin{align}\label{PE}\!\!
{\Phi_D} \lesssim  \left( \frac{M}{M_{\odot}}\right)\left(\frac{\text{kpc}}{R}\right)^2 \left(\frac{\text{keV}}{E_D}\right) \times10^3  /(\text{cm}^{2}\cdot\text{s} ).
\end{align}
In Figure \ref{figFlux}, we make the contour plot of dark flux $\Phi_D$ as a function of locations in the disk $R/R_s$ and black hole masses $M/M_{\odot}$, 
which will be helpful to estimate the benchmark parameters  below.

%which will be used in the estimations below.
%which will be used in the following estimations.
% ${\mathcal{E}_E}$
%&\equiv {\mathcal{E}_D}/E_D=\frac{L_D}{4\pi R^2 E_D} \nn\\&

\subsection{Possible signals and benchmark parameters} %Possible signals for direct detections} 
%Axion and dark photon, dark matter
 %and the flux in the line was around the order of  $10^{-5}$ photons/ (cm$^2\cdot$ s$\cdot $ deg $^2$).One key parameter that one need to check is whether the flux of AGN can reach that excess of Xenon1T \cite{Buch:2020mrg}.

%{\cm [ref to Page 9-11 of the slide Newton2020 Xenon].
%In order to explain the Xenon1T excess at the keV scale,}

 %  This flux can be a benchmark point for the dark matter fluxes from the black holes.  %Notice that this is two orders of magnitude higher than the dark matter in an NFW profile is $\sim $ \cite{Navarro:1995iw} from galactic center.

Below we discuss the light dark matter candidates that can be produced around black holes and estimate their flux. Taking into account of known AGN or BHB sources, we find that the flux is too low to explain the {\XT} excess.
However, it is interesting to estimate the benchmark parameter of the flux, which may be detected in the future experiments. 

Before that, we summarize the general result on the detection.
One possibility is to consider the electron recoil, and the total number of signal events is given by
$N_s=\Phi_i \times \sigma_{ie} \times T_X C_X$.
%\begin{align} N_s=\Phi_i \times \sigma_{ie} \times T_X C_X,\quad C_X=Z_e n_X V_X. \end{align}
Here $ C_X=Z_e n_X V_X$ and $\Phi_i$ is the incoming flux with dimension [cm$^{-2}$ s$^{-1}$],
$\sigma_e$ is the cross section with dimension [cm$^2$] and $T_X$ is the operation time.
$C_X$ is the total number of effective electrons in the detector, which can be calculated by the product of the effective number $Z_e$ of electrons in each Xenon atom that undergo recoils, the number density of Xenon atoms $n_X$  and the total fiducial volume $V_X$.
On the {\XT} excess with $N_s\sim O(100)$,
e.g. in \cite{Fornal:2020npv}, the required product of the number flux and cross section has been found to be
$ %\begin{align}\label{CS}
\Phi_i \times \sigma_{i} \simeq O(10^{-35})/s.
$ %\end{align}
In the following, we will use this relation to estimate the required flux and then the distance to the sources with \eqref{PE}, considering the accreting black holes as the sources.
%{\cm For example, the flux from Planet 9 can reach $10^{-28}/s$, which is a good candidate to consider \cite{Scholtz:2019csj} .}

\vspace{10pt}
{\it 1. Axion/Axion-Like particles(ALP)}.
One possible explanation of the claimed {\XT} excess is the solar axion/ALP. It requires the coupling to electrons without/with photon coupling through the inverse Primakoff process \cite{Gao:2020wer},
$a+Ze\to \gamma+ Ze $ or inverse Compton scattering $a+e\to e +\gamma$, with the couplings in the Lagrangian
 \begin{align}
 \mathcal{L}_{a }=- \frac{g_{a\gamma}}{4} a \,F_{\mu\nu}\tilde{F}^{\mu\nu}- \frac{g_{ae}}{2}\frac{\partial_\mu a}{m_e}\, \psi_e\gamma^\mu \gamma_5\psi_e.
\end{align}
Since the solar axion luminosity is bounded by stellar cooling, the couplings have been constrained as $g_{a\gamma}{<}10^{-10}\text{GeV}^{-1}$ or $g_{ae}{<}10^{-12}$ \cite{Redondo:2013wwa}.
%{\cm  if adding $f_a$, $f_a$ is conventional chosen as $10^{10}$ GeV, so just plz keep the current convention}
The kinetic energy of axion/ALP considered for Xenon here is 2-4 keV with the axion mass $m_a \ll 1 $eV. However, this explanation is in tension with astrophysical stellar cooling constraints \cite{DiLuzio:2020jjp},  because the solar axion production is well studied.

What's more, the couplings bound from the stellar cooling constraints also lead to the cross sections bound.
%Considering \eqref{CS},
The minimum required number flux density of the  incoming axions for the {\XT} excess is around
 ${ \Phi_a} \gtrsim 10^{11}/(\text{cm}^{2}\cdot\text{s})$.
%\begin{align}{ \Phi_a} \gtrsim 10^{11}/(\text{cm}^{2}\cdot\text{s}) .  \end{align}
 %which can explain the Xenon excess.
Here instead, it is possible to consider accreting black holes as the sources. If we consider the keV axion in the dark flux bound by \eqref{PE}, with the help of Table \ref{tableBH1}, one can obtain the minimum distance of the three types of black holes. It is tempting to explain the Xenon 1T axion with the black hole sources rather than the Sun, provided that the thermal plasma around the observed black hole is at the keV temperature.

\vspace{10pt}
{\it 2. Light Dark Photon}.
%The solar dark photon is a little soft for the excess as in \cite{An:2020bxd}.
The dark photon production in the Sun is mostly from dark Higgstrahlung and visible photon oscillations with real or virtue photons.
The dark photon and dark fermion sector is given by the Lagrangian density
%${\gamma'} + {\gamma'}   \to \gamma + \gamma$ (what is this? this is not the required process) with the mix coupling
\begin{align}\label{chiF}
 \mathcal{L}_{{\gamma'}}=&\bar{\chi}(i\slashed{D} +m_{\chi})\chi -\frac{1}{4}( {F}'_{\mu\nu})^2 %\nn \\&
+ \frac{m_ { {\gamma'}}^2}{2}  {A'}^2 +\frac{{\epsilon}}{2}   {F}'_{\mu\nu}F^{\mu\nu} .
\end{align}
The typical solar flux is
$\Phi_{{\gamma'}}\lesssim 10^{10} /(\text{cm}^{2}\cdot\text{s}  )$ for mixing ${{\epsilon}} \sim O(10^{-14})$  \cite{An:2013yua}.
%(The Lagrangian is amended by the Higgs-strahlung processes into dark photon pairs lead to a stellar energy loss constraints \cite{An:2013yua}.) {\cm This statement is not correct}
It was recently pointed out in \cite{An:2020bxd} that solar emission of dark photons is a little soft to fit well with the recent {\XT} excess.

Here if we replace the Sun by other sources such as two types of accreting black holes in our Table \ref{tableBH1}, with different couplings, the flux aimed at the {\XT} excess received on Earth is $ { \Phi_{{\gamma'}}} \gtrsim 10^{11}/(\text{cm}^{2}\cdot\text{s})$.
%\begin{align}{ \Phi_{{\gamma'}}} \gtrsim 10^{11}/(\text{cm}^{2}\cdot\text{s}) .\end{align}
Considering a few known keV X-ray sources combined in \cite{Ozel:2010su,Barack:2018yly}, 
and with the help of Figure \ref{figFlux} or equation \eqref{PE}, one can summarize in Table \ref{tableDM1} below with $\sim$kpc distance black holes with the assumed dark luminosity $L_D\sim L_E$ \cite{Ozel:2010su,Barack:2018yly} . However, the maximal intensity of the flux $\Phi_D$ is around $10^{3}/(\text{cm}^{2}\cdot\text{s})$, which is a few orders lower for the Xenon1T excess, but may be considered for the future astrophysical observation.
% experiments. 
% with the luminosity

Here we would also like to comment on the X-ray telescope constraints, similar to the discussions in \cite{Dessert:2019sgw}. In general, these benchmark points in Table \ref{tableBH1} do not contradict the X-ray observation, since we basically assume the dark flux/axion flux is converted from the observed photons in X-ray. So the dark flux is around the same order as the X-ray observations and does not exceed the observed limits.
In the other words, the explanation that the Xenon1T signals come from the known X-ray sources of AGN and BHB conversions is basically excluded. In Figure \ref{figXenon}, we use the green dashed line to denote the bound from the X-ray background in \cite{Ballesteros:2019exr}.
In the next section, we will consider the dark fluxes from the accreting black holes directly.

 \begin{table}[th]
\begin{tabular}{ |c|c|c|c }
\hline%/$(\text{cm}^{2}\cdot\text{s}\cdot \text{keV})$
DM Types   & Axion or ALP($a$) & Dark Photons ($\gamma'$)       \\
\hline
Masse (m/keV) & $ <10^{-3}$ & $<10^{-2}$    \\
\hline
Energy(E/keV) & $ \sim 3$ & $\sim 3$    \\
\hline
Flux($\Phi_D$/({cm}$^{-2} \cdot$ {s}$^{-1}$))   & $\sim 10^{3}$   & $\sim 10^{3}$    \\
\hline
Couplings & $g_{a\gamma} \sim 10^{-10} \text{GeV}^{-1}$ & ${\epsilon} \sim 10^{-14}$  \\
 &  and  ~$g_{ae} \sim 10^{-12}~$~~ &     \\
\hline
%Distances (R/kpc) & $< 10^{-4}$  & $< 10^{-4}$      \\
%\hline
\end{tabular}
\caption{ {\cm Here we list two benchmark points with some known black hole binary sources combined in \cite{Ozel:2010su,Barack:2018yly}. We assume the dark particle luminosity of the host black hole is comparable to the visible luminosity $L_D\sim L_E$, and those sources are around kpc distance from the earth.
}\label{tableDM1} }
\end{table}

%\vspace{10pt} %{\it 3. Dark Photon Dark Matter.}

\section{Dark Matter Flux and Black Hole Dynamics}%{Dark Matter Flux from Black Holes} 
%\section{dark matter with self-interactions from Black Holes} % from Accreting Black holes  } %Axion and dark photon,  source
\label{Flux}
 {\cm 
In this section, we will discuss how heavier dark matter may accumulate, and accelerate around the black hole,as well as the benchmark parameters compared with the Xenon1T experiment.}
%Especially for some dark particles, the creation and acceleration do not necessarily require the standard model couplings.
%{\cm
%We will also discuss the acceleration of dark matter particles by the black hole accretion.}
%As also commented in \cite{Kannike:2020agf},
%{\cn (YL: tba: have to consider dark photon? or dark BZ? )}

%\label{Flux1}
%\subsection{ Emission from AGNs and X-ray Binaries}

%Detailed simulation of this process will be an interesting future topic. In the next subsection, we estimate the orders of magnitude of the flux to compare with the Xenon excess.

\vspace{10pt}
%{\it 3. Boosted Dark Matter and Black Hole - Dark Star Binary.}
%\subsection{ Self-interacting Dark Matter} % Self-interacting 
\subsection{Dark luminosity and accreting black hole}
Now we consider a different mechanism to produce an outflow of dark matter particles.
 The heavier non-relativistic  MeV-GeV dark matter cannot be thermally produced in the plasma. They can however get boosted around the event horizon or captured/accreted by the black hole.  Especially the black holes have been considered to act like particle accelerators to boost the particles to arbitrary high energy and collide \cite{Banados:2009pr,Jacobson:2009zg}.
 However, for the traditional collision-less cold dark matter, the dark matter particles can have a larger speed near the black hole, but that does not mean they can escape. If we assume the collision can happened, then some of the products may escape the near horizon region of black holes. That means that we need to consider self-interacting dark matter in order to account for a dark flux \cite{Tulin:2017ara}.
 {\cm 
Another possibility is the charged DM model with the dark electromagnetism (mediated by the dark photon) as discussed in the lagrangian \eqref{chiF}.
One may be able to invoke shocks or magnetic reconnection processes to accelerate particles via the first-order Fermi acceleration mechanism. All these processes require charged dark matter particles.
}

At the galaxies scale,  the self-interacting dark matter model is introduced to solve the  problems in small scale structures \cite{Tulin:2017ara},
such as the core-cusp problem, the rotation curve's diversity problem, the missing satellites problem, etc. The double-disk dark matter is also motivated \cite{Fan:2013yva, Fan:2013tia}, such that the accretion disk model with viscous hydrodynamical approximation can be applied. For example in \cite{Ackerman:2008kmp}, the hypothetical dark photon was proposed to mediate the interactions of dark matter particles through \eqref{chiF}.
%$\mathcal{L}_{\chi\gamma'}=\bar{\chi}(i\slashed{D} +m_{\chi})\chi-\frac{1}{4} {F'}^2$. 
Neither of them interacts much with the standard model particles.

%\begin{align}\label{dark} \mathcal{L}_{\chi\gamma'}=\bar{\chi}(i\slashed{D} +m_{\chi})\chi-\frac{1}{4} {F'}^2. \end{align}
%such that the dark matter particles can be annihilated into the dark radiation $\chi\chi\to \gamma'\gamma'$.
%Those particles
%In the next subsection, we will estimate the dark luminosity for the dark radiation from self-interacting dark matter.

% can be produced from $\chi\chi\to \gamma'\gamma'$, , and  is the cross section between dark photon and dark matter particles
%recall the derivation of Eddington luminosity in \eqref{EL},

It is interesting to estimate the dark luminosity following the similar derivation of Eddington luminosity in \eqref{EL}. 
{\cm
In \cite{Outmazgine:2018orx}, it has been found that that accretion of dissipative dark matter 
onto AGN contributes to the growth rate of the black hole.
Here we assume the dark matter particles are scattered upon the dark photons via the cross section $\sigma_{\chi\gamma'}$ due to the interaction in \eqref{chiF}, }
we can have the dark luminosity % $L_{\chi}  or,
\begin{align}\label{DL}
L_{\chi}=\frac{4\pi G M c}{\kappa'_\chi}  =L_E \frac{\kappa_E}{\kappa'_\chi},
\end{align}
where $\kappa'_\chi\equiv  {\sigma_{\chi\gamma'}}/{m_\chi}$.
For ionized hydrogen, one has
$ \kappa_E =\sigma _{\rm {T}}/m_{\rm {p}} \sim 0.3  \text{cm}^2/\text{g}$.
%For the case of $m_\chi\sim$keV, $\frac{m_p}{m_\chi}\sim 10^6$, but $\frac{\sigma_\chi}{\sigma_T}=\frac{g_{ae}^2}{\alpha^2}\sim 10^{-20}$.
From numerical simulations of merging cluster galaxies \cite{Randall:2007ph, Tulin:2017ara},
one bound of the self-interaction cross-section of dark matter particles is $\frac{\sigma_{\chi\chi}}{m_\chi} \lesssim 1 \text{cm}^2/\text{g}$. Here the cross sections $\sigma_{\chi\chi}$ and $\sigma_{\chi\gamma'}$ are both related to the dark photon fine-structure constant ${\alpha'}$, which is bounded by the relic abundance  \cite{Ackerman:2008kmp}.  Hence, it is reasonable to assume $\kappa'_\chi \sim \kappa_E$, and we can have the dark luminosity as bright as the Eddington luminosity $L_{\chi}\sim  L_E$.   If considering the case of the self-interacting dark matter without dark radiation, the gravitation could be balanced by the pressure of the dark particle gas $\kappa_{\chi\chi}\equiv  {\sigma_{\chi\chi}}/{m_\chi}$, then again we can reach $L_{\chi}\sim  L_E$. % $\kappa_{\chi\chi} \lesssim 1 \text{cm}^2/\text{g}$ and

\begin{figure}[h]
\centering
%\raggedright
\includegraphics[scale=0.3]{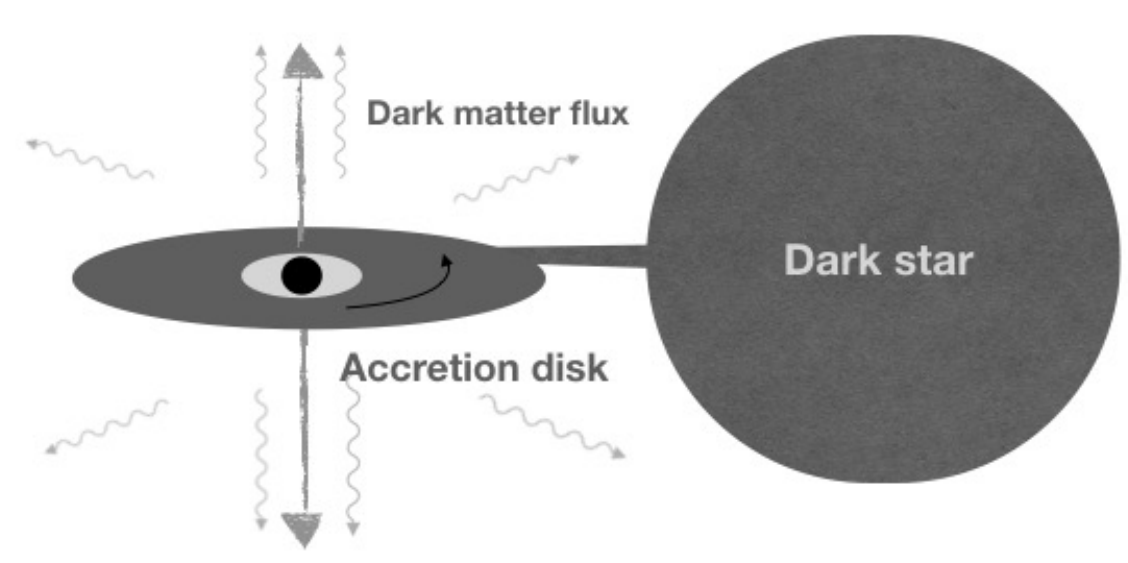}\qquad \\
\caption{
The schematic diagram of black hole and dark star(or axion clump) binary system. 
If dark matter particles have enough self-interactions,  they can clump together to form a dark star. If such dark star happens to be accreted to one black hole, it is possible to form the dark matter disc and dark matter flux.
\label{figBH}}
\end{figure}

For the much lighter axion, the self interaction is very small \cite{Desjacques:2017fmf}.
%{\cm (YL:  constrain mass of axion from here? )}
%The cross section $\sigma_{a\gamma'}\simeq \sigma_{a\gamma} \sim g_{a\gamma}^4 E^2  \sim 10^{-79}$cm$^2$ with $E\sim$ keV.
% to reach $\kappa'_a\sim\kappa_E\sim 0.3  \text{cm}^2/\text{g}$, we obtain $m_a\sim  10^{-79}\text{ g} \sim 10^{-36}\text{eV}$.
% which is consistent with the QCD axion mass regime.
If we take the cross section $\sigma_{a\gamma'}\simeq \sigma_{a\gamma}$ and usual QCD axion at $\mu$eV scale,
 then $\kappa'_a\equiv \frac{\sigma_{a\gamma'}}{m_a}$ will be many orders of magnitude smaller than $\kappa_E$,
which means that the luminosity $L_{a}=L_E \frac{\kappa_E}{\kappa'_a}$ can be much higher.
%to be $R \lesssim \left({m_a}/10^{-4}\text{eV}\right)^{1/2}$kpc (need to change).
%Although the distances in Table \ref{tableDM1} bound by the visible Eddington luminosity can be relaxed,
However, this kind of system is usually unstable and may lead to bursts of photons or axions after a short time. This process may be an interesting source to search for axions and will be interested in further exploration. %ed this in some future work.Alternatively, 
For example, one may consider a possible black hole dark star binary system which is depicted in Figure \ref{figBH}. If dark
matter particles have enough self-interactions, it can clump together to form a star. We can see the examples of e.g. boson stars, axion stars or moduli stars \cite{Braaten:2019knj,Sun:2020gem}. If such a dark star happens to be accreted by a black hole, a dark disk may form around the black hole. Similar to an X-ray binary, this object, if exists, could be a power source of dark matter flux, without necessarily being observed in the X-ray band, such that it could be quite close to Earth. Since these systems are mostly invisible, it is possible for the black hole - axion star binaries to produce the required relativistic flux of axions with self-interaction. More detailed modeling of black hole dynamic processes is required for a similar production rate calculation around the black holes.
In the following, we list a few mechanisms to produce the dark matter flux.

%Similarly, other models utilizing the vector portal can fit the excess, including those with operators that directly couple the dark photon field strength to electron spin.
%\begin{align}
%\frac{dN}{dE}   \sim 10^{10}/(\text{cm}^{2}\cdot\text{s}\cdot \text{keV} )%\left(\frac{\epsilon}{10^{-15}}\right)
  %\end{align}

%and the black hole accretion can be one way thermaliz it to be keV scale.

\vspace{10pt}
\subsection{Boosting dark matter by black holes}
% for Xenon1T

There are a few interesting circular orbits very close to the event horizons of the black holes.
The photon sphere near a black hole is where photons travel in circular orbits.
For a Schwarzschild black hole with the event horizon located at $\Rs=\frac{2GM}{c^2}$, the photon sphere is located at $3\Rs/2$. And the innermost stable circular orbit of the massive particles is located at $3 \Rs$. For a rotating black hole, it has corrections due to the rotational effects \cite{Bardeen:1972fi} which can be diagnosed using BH images obtained from EHT (see e.g.~\cite{Akiyama:2019cqa}). In general, for a rotating and charged black hole with angular momentum  $J$ and charge $Q$, the outer event horizon is $R_{+}=\Rs(1+\sqrt{1-a^2-q^2})/2$, where $a=J/M$ and $q=Q/M$. Here when $a^2+q^2=1$, $R_{+}=\Rs/2$ is the extremal radius. Normally $a^2+q^2<1$ is required by the cosmic censorship hypothesis to avoid naked singularities. The Hawking temperature sourced by the quantum radiation of a Schwarzschild black hole is around
$T_H \simeq 6\times 10^ {-8} K \Big(\frac{M_{\odot}}{M_{BH}}\Big)$  \cite{Hawking:1974rv}.
For stellar-mass black holes around 3-10$M_{\odot}$, one has $T_H \sim 10^ {-8} K$.
For super massive black holes ($\gtrsim 10^6 M_{\odot}$), one has $T_H \lesssim 10^ {-13} K$.
Both are much lower than the observed temperature of the cosmic microwave background temperature 2.7K.
However, some black holes are surrounded by very hot coronal thermal plasmas and accretion disks, with the inner circle located around ISCO $R_{ {I}}=3\Rs$.  For the primordial black holes with much smaller masses, the Hawking temperature can be as high as keV$\sim 10^7 K$, and we leave the discussion of this case to the next section.

 %much smaller mass 

%Hereof the black hole.
%For the Kerr or Reissner-Nordstrom black hole.Kerr–Newman metric

The heavier boosted dark matter from MeV to GeV mass with around 0.1c can also explain the excess as in \cite{Kannike:2020agf, Fornal:2020npv} with a number flux of $\Phi_\chi\sim 10^{-6} /(\text{cm}^{2}\cdot\text{s} )$ in an NFW profile \cite{Navarro:1995iw} from the Galactic center to produce Xenon excess considering elastic scattering between the dark matter particle and  an
electron $\chi + e \to \chi + e'$.
For the non-relativistic case, the transferred recoil energy 
%is $E_R\simeq 2m_e v_\chi(v_\chi-v_e)$, for the case of $m_{\chi}\gg m_e$.
\begin{align} E_R\simeq 2m_e v_\chi(v_\chi-v_e), ~\text{for}~m_{\chi}\gg m_e. \label{recoil}
 \end{align}
Considering the mass of electron as $m_e\simeq 511$keV and $v_e\sim 10 \alpha_e\simeq 10/137$ as in \cite{ Kannike:2020agf}, for the {\XT} excess between 2-4 keV, a required benchmark velocity of the heavy dark matter particles is $v_\chi\sim 0.1c$, with the boost factor $\gamma_{0.1}=\left(1-v_\chi^2/c^2\right)^{-1/2}\simeq1.005$.
The flux has been calculated to be  $\Phi_{\chi}\simeq 10^{-6}/(\text{cm}^{2}\cdot\text{s}) \left(\frac{3\times10^{-29} \text{cm}^2}{ \sigma_{e\chi}} \right)$,
% \begin{align} \Phi_{\chi}\simeq 10^{-6}/(\text{cm}^{2}\cdot\text{s}) \left(\frac{3\times10^{-29} \text{cm}^2}{ \sigma_{e\chi}} \right), \end{align}
where $\sigma_{e\chi}$ is the required cross section
and here we just use it as the benchmark flux for our dark sources \cite{Kannike:2020agf, Fornal:2020npv}.

Black holes can be good accelerators to accrete and boost heavy dark matter particles. For the boosted dark matter to escape and reach the benchmark speed of 0.1c at the Xenon detector, we need to assume extra ingredients. One may consider the semi-annihilation dark matter $\chi'\chi'\to \chi \phi$, where $\chi'$ and $\chi$ are the heavy DM particles with the same mass, and $\phi$ is a light particle coupled to the SM sector \cite{DEramo:2010keq}. The boost factor of the final states turns out to be ${\gamma}_{{\chi}}=
\gamma_{i}+\frac{1-m_\phi^2/m_\chi^2}{4\gamma_{i}}\leqslant \gamma_{i}+\frac{1}{4\gamma_{i}}$, where $\gamma_{i}$ is the boost factor of $\chi'$ in the center of mass frame. When $\gamma_{i}=1$, we reach the limit $\gamma_{\chi}=1.25$ and  $v_{\chi}=0.6 c$. For the two-component DM model $\chi_A \bar{\chi}_A\to\chi_B\bar{\chi}_B$ studied in \cite{Fornal:2020npv} with initial boost factor $\gamma_{A}$,
the boost factor for $\chi_B$ is simply given by $\gamma_B=\gamma_A m_A/m_B$ \cite{Boehm:2003ha}, which can go beyond $0.6c$ easily.
Thus, for both cases, the velocities of the produced dark matter particles $\chi_B$ can reach or go beyond the escape velocity outside ISCO. The dark accretion disk also provides a higher density environment to enhance semi-annihilation and collision rates, and the products could then escape. Alternatively, we need to consider the charged DM model with the dark electromagnetism sector \eqref{chiF}, 
%as discussed in section \ref{Flux1},
 such that it is possible to reach a larger flux as described around the dark luminosity in eq. \eqref{DL}.

%  semi-annihilation to occur and final state DM particles to escape
%when $m_{\phi}\ll m_\chi$, and it is larger than $v_{\text{}}=0.5c$.
%To see this, we consider the semi-annihilations around the Schwarzschild black hole.
%Thus, once setting $\gamma_B>\gamma_{0.5}\simeq 1.155$, it's also reasonable to put it surrounding the black holes.
%Taking into account the gravitational redshift of the final state particle ${\chi'}$ that emits radially, we have $\gamma_{0.1}m c^2=  \gamma_{{\chi}} m c^2(1- {\Rs}/{R})^{1/2}$, that lead to $R=\Rs/(1- {\gamma_\infty^2}/{\gamma_{\bar{\chi}}^2}) \gtrsim 2.83\Rs$.
 %  \left(\frac{N_{s}}{100}\right).
%However, the collisional Penrose processes are only effective for the  in rapidly rotating black hole
 %The dark matter particles may collide with each others through self-interactions  however,  it's not easy for them to escape the black holes

\begin{table}[th]
\begin{tabular}{ |c|c|c|c }
\hline%/$(\text{cm}^{2}\cdot\text{s}\cdot \text{keV})$
DM Types    & Boosted DM($\chi$)& DPDM($\gamma'$) \\
\hline
Masse (m/keV)  & $\sim 10^{3}-10^6$ 
& $ \sim3$ \\
\hline
Velocity(v/c)   & $\sim 10^{-1} $ 
& $\sim10^{-3}$  \\
\hline
Flux($\Phi$/({cm}$^{-2} \cdot$ {s}$^{-1}$))     &  $\gtrsim 10^{-6}$ 
& $\gtrsim 10^{12}$  \\
\hline
Parameters  & $\sigma_{ e\chi} \sim 10^{-29}$cm$^2$ 
&    ${\epsilon} \lesssim 10^{-15}$\\
\hline
%Required  & No requirement &   \\
%BH booster 
Benchmark (R/kpc)      & $ < 10^{1} - 10^{3}$   & --  \\
%Distances (R/kpc)  &narrow parameter choices) &  \\
 \hline 
\end{tabular}
\caption{
Here we list the other two benchmark points for the {\XT} explanations with black hole sources.
The parameter choices are based on \cite{An:2020bxd} and \cite{Fornal:2020npv}.
 We assume the dark particle luminosity is $L_D\simeq 10^{38} \text{erg/s}$.
 The last row shows the benchmark distances that can produce the Xenon1T excess. 
 For the dark photon dark matter(DPDM), there is no requirement. The BHs help to increase the flux $\Phi_{\gamma'}$ and thus relax the benchmark choice of mixing parameter $\epsilon$.
 \label{tableDM2}}
\end{table}

Here we again assume that the dark luminosity is characterized by the Eddington luminosity. In Table \ref{tableDM2}, we list heavier dark matter candidates, such as the dark photon dark matter \cite{Nelson:2011sf}, where the dark photon is produced non-thermally e.g. from inflation decay and acting like cold dark mater, or boosted dark matter with a black hole source. The parameter choices are from \cite{An:2020bxd} and \cite{Fornal:2020npv}, and we assume that the dark particle luminosity of the host black hole is $L_D\simeq 10^{38} \text{erg/s}$.

\begin{figure}[h]
\centering
%\raggedright
\includegraphics[scale=0.5]{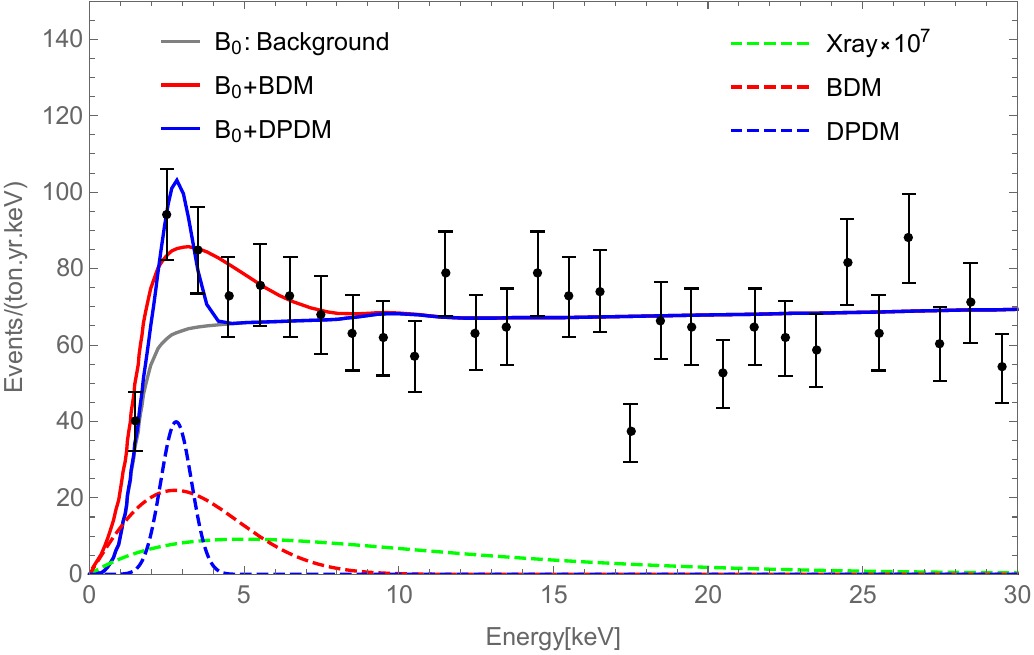}\qquad \\
\caption{
The data points and background curve $B_0$ are reproduced from the Xenon1T data and the background model in \cite{XENON:2020rca}. The red dashed line is produced from the boosted dark matter (BSD) model accelerated by the black holes. This line is schematically produced as in \cite{Fornal:2020npv}, with $v_\chi\simeq0.1c$ and other benchmark parameters in Table \ref{tableDM2}. Here we smear the recoil energy in equation \eqref{recoil} with the Rayleigh distribution, considering the detector resolution. The blue dashed line denotes the signals of the dark photon dark matter(DPDM) with $m_{A'}=2.8$keV with resolution of 1keV as in \cite{An:2020bxd} and other benchmark parameters in Table \ref{tableDM2}.
The green dashed line denotes the axion flux bounded by the X-ray background in \cite{Ballesteros:2019exr}. The event number is shown in the plot after being multiplied by the factor of $10^7$.
\label{figXenon}}
\end{figure}

There are also a lot of discussions on the cosmological origin of the dark photons (see e.g. \cite{Salehian:2020dsf}), similar to the dark photon dark matter model. The mass of the dark photon dark matter is generally heavier than the dark photon case we discussed in section \ref{Xray}. It was found in \cite{An:2020bxd} that the local dark photon dark matter with mixing angles ${\epsilon}\sim O(10^{-16})$ can fit the recent Xenon1T excess well and satisfy the astrophysical constraints. Here one can find for the local 3 keV dark photon dark matter with the density $\rho_{{\gamma}'}\simeq 0.3 \text{GeV/cm}^3$ and $10^{-3}$c, the flux can be as high as $\Phi_{\gamma'}\sim 10^{12}/(\text{cm}^{2}\cdot\text{s} )$.
The intrinsic mass $m_{\gamma'}\sim 3$keV is built into the model \cite{An:2020bxd}.
In fact, if we consider the black hole sources, the flux can be enhanced, and mixing parameters ${\epsilon}$ and $m_{\gamma'}$ can then be relaxed accordingly.
In Figure \ref{figXenon}, we plot the Xenon1T signals, background and the predicted signals from two scenarios discussed in Table \ref{tableDM2}. 

With the benchmark parameters and Figure \ref{figXenon}, we would also like to comment on the dark matter constraints from direct detections.
For the boosted dark matter case, the black hole accumulation and boosting mechanism does not depend on the DM-nucleon scattering cross section \cite{Fornal:2020npv}, different from solar capture mechanism, therefore the current cross section $\sigma_{ e\chi} \sim 10^{-29}$ cm$^2$ 
does not subject to the other DM direct detection experiments constraints in general.

For the dark photon dark matter in \cite{An:2020bxd},  the star cooling constraints are stronger than other direct detection limits e.g. XENON100~\cite{XENON100:2014csq}, XENON1T~\cite{XENON:2019gfn}, SuperCDMS~\cite{SuperCDMS:2019jxx}, CDEX\cite{CDEX:2019isc}, SENSEI~\cite{SENSEI:2019ibb,Tiffenberg:2017aac,SENSEI:2020dpa} and CRESST-III~\cite{CRESST:2015txj}. % in our two benchmark scenarios here.  in this dark matter mass range.
But in our case, the flux of dark photon dark matter is enhanced by black holes with $\Phi_{\gamma'}\gtrsim 10^{12}/(\text{cm}^{2}\cdot\text{s} )$ around $m_A \sim 2.8$keV. The mixing parameter can then be relaxed $\epsilon \lesssim 10^{-15}$. 
Thus, our benchmark parameters choice in Table \ref{tableDM2} is safe from direct detection constraints.

\vspace{10pt}
\subsection{Dark superradiance of black holes } %Other Alternatives on dark radiation} %Discussion on 
\label{others}

On the superradiance of black hole, one kind of interesting objects is the gravitationally bound ``atoms" with light bosons (e.g. axions) or vectors \cite{Arvanitaki:2009fg}. Similar to the hydrogen atoms,  those light particles occupy different energy levels of the rotating black holes. Although the kinetic energy in the superradiance clouds is non-relativistic, these accumulated light particles can be accelerated through different energy extraction processes around the black hole.
 Notice that in gravitational atoms, when the Compton wavelength of dark particle is comparable to the size of the black hole,  the number of dark particles can grow exponentially from extracting energy of the rotating black holes.
 When the attractive self-interactions become stronger than the gravitational binding energy, the cloud collapses and the explosion induces an outflow \cite{Arvanitaki:2009fg}.
If considering the decay of axion through $a\to \gamma\gamma$, it has been shown in \cite{Rosa:2017ury,Sun:2020gem} that the $\mu$eV axion clouds around $10^{-5}M_{\odot}$ PBHs may induce fast radio bursts (FRBs), and the luminosity can be as high as $10^{39}$erg/ms, more than enough to power observed FRBs.
% Though the time scale of FRBs is around milliseconds, the total energy released is comparable to the energy released by the Sun in one day.

It is also interesting to consider the bursts of dark sectors with the similar mechanism.
If considering the channel of $\phi \to\gamma'\gamma'$, the dark photon at 3.5 keV requires $m_\phi\simeq7$keV.
Considering the reduced Compton wavelength $\lambda_\phi= {\hbar}/{(m_\phi c)}$, the dimensionless number describing the dark matter superradiance cloud is
$ %\begin{align}
\alpha_s \equiv\frac{R_s}{\lambda_\phi}
\simeq  0.1\times\left(\frac{M_{BH}}{10^{-15}M_{\odot}}\right)\left(\frac{m_{\phi}}{7\, \text{keV}}\right).
$ %\end{align}
If we take the typical value of $\alpha_s$ as the order of 0.1, it is interesting to see that the black holes 
with mass $10^{-15}M_{\odot}$ also have the Hawking temperature $T_H \sim$ keV. % with \eqref{Hawking} . 
{ Both effects are characterized by the quantum wavelength of the black hole size. } To see this, consider Hawking radiation with temperature
\begin{align}\label{Hawking}
T_{BH} = \frac{\hbar }{4\pi  k_B}\frac{c}{\Rs} \simeq 5.3 \text{keV} \left(\frac{10^{-15}M_{\odot}}{M_{BH}}\right).
\end{align}
Requiring $k_B T_{BH} =3.5  \text{keV}\simeq 4\times 10^7 K$, we obtain  $M\simeq 1.5\times 10^{-15} M_{\odot} $,
which is in the mass region of primordial black holes \cite{Sasaki:2018dmp}.
The life time for such PBHs to evaporate is around
$\tau_{BH}\simeq 10^{22} \text{years}(\frac{M_{BH}}{10^{-15}M_{\odot}})^3$, which is much longer than the age of the universe at $10^{10}$ years. If the particles in Hawking radiation of PBHs are mainly in the dark sectors \cite{Hooper:2019gtx},
 dark radiations at the keV energy scale could also be one candidate source for dark fluxes.

The  Bekenstein-Hawking luminosity of a Schwarzschild black hole is given by $A_s \sigma_s T_{BH}^4 $, with the area of horizon $A_s=4\pi R_s^2$. It is calculated to be
$
L_{BH} =\frac{\hbar }{15\pi}\frac{c^2}{(16\Rs)^2} \simeq 2\times10^{-25} L_{\odot}\left(\frac{10^{-15}M_{\odot}}{M_{BH}}\right)^2.
$
Thus, even for these $\sim 10^{-15} M_{\odot}$ PBHs, the luminosity of Hawking radiation $L_{BH}\sim 10^{-25}L_{\odot} $ is quite small compared to the solar luminosity. 
%Due to the small luminosity, one might not need to consider them here for {\XT}.
To estimate the total flux, one needs to consider the integrated luminosity with the distribution of PBHs. 
% the density distributions of the axion like particles from primordial black holes produced in the early universe.
The related X-ray constraints have already been discussed in the PBHs as the dark matter scenarios with the Hawking radiation in \cite{Coogan:2020tuf,Laha:2020ivk,Ballesteros:2019exr}.
It is also interesting to consider whether we can find axion signals from PBHs in X-ray telescopes. See more recent discussions in \cite{Schiavone:2021imu}.
The axion flux converted from the known X-ray observation is plotted as the dashed green line in Figure \ref{figXenon}, which is a few orders lower than the Xenon1T excess.

%\vspace{10pt}
\section{Discussion and Summary}
\label{Con}

In this study, we propose that {accreting} black holes could be the natural sources to produce the keV-order light dark matter and boost the heavier dark matter. We discuss several mechanisms to produce, accumulate and accelerate the dark photon, axion, even heavier dark matter around the black holes, motivated by the fast dark matter explanations of {\XT} excess, e.g. \cite{Smirnov:2020zwf}-\cite{Sakstein:2020axg}.

The X-ray black hole binares with stellar masses, or AGNs in the super massive black holes can be possible hosts (see also a recent event of an intermediate mass black hole \cite{Abbott:2020tfl}). However, the fluxes observed in AGNs and BHBs are many orders of magnitude smaller than the flux required for the {\XT} excess %due to the distance
since their distances to Earth are $> 1$ kpc.
%  to the Earth. 
Although the black hole corona and disk productions of axions and dark photons require some detailed modeling of accretion processes,  the observed X-ray emissions provide benchmark points around the 1-8 keV range.  There are many bright X-ray sources in the sky that originate from accreting or spindown-powered neutron stars. Supernova remnants, Gamma ray bursts (GRB) afterglows can also produce bright X-rays although the density in the emission region is extremely low and the radiation process is synchrotron radiation. These systems could also be interesting subjects of study in the future.

{\cm One viable mechanism for the current Xenon1T excess is black holes as the dark matter boosters. We discuss and calculate the boosted factor and dark luminosity here. The black hole accretion can also accumulate dark matter to achieve a higher dark matter local density. }
The black hole candidates such as primordial ones (e.g. the smaller ``Planet 9" with $10^{-5}M_{\odot}$) with closer locations may also be the sources to boost dark matter  \cite{Scholtz:2019csj,Witten:2020ifl}.  The other possible region is the micro PBHs  around $10^{-15}M_{\odot}$ which can  emit X-rays through thermal radiation. Although the luminosity is quite small, it is plausible to assume the radiations partly belong to the dark sector.  
%. The Hawking radiations are theoretically calculated based on the semi-classical approximation and
%%For example the so-called%% ``Planet 9" with the assumed mass $\sim 10^{-5}M_{\odot}$ \cite{Scholtz:2019csj,Witten:2020ifl}, is quite close to the earth, and it is a very good candidate to consider here.

 The dark cosmic fluxes created around black holes can be interesting and less explored targets for direct detections, in light of the multi-messenger detection experiments for black holes such as gravitational waves and EHT. The constraints are also less severe compared to the stellar productions. Especially with the help of the multi-messenger detections,
we hope the dark processes around black holes can be understood and modeled better in the near future.

%%
%%
% move PBH to the next paper
%%
%%

\vspace{10pt}
\section*{ Acknowledgments}
%\vspace{-10pt}
%\small %\footnotesize
This work was supported in part by the National Natural Science Foundation of China (Grants No. 12105013, 12005255, 11690022, 11947302, 11991052, 11690022, 11821505 and 11851302), the Key Research Program of the Chinese Academy of Sciences (CAS Grant No. XDPB15), the Strategic Priority Research Program of CAS(Grant No. XDB23030100), and the Key Research Program of Frontier Sciences of CAS.
This work was also supported by MIUR in Italy under Contract(No. PRIN 2015P5SBHT) and ERC Ideas Advanced Grant (No. 267985) \textquotedblleft DaMeSyFla", Grant-in-Aid for JSPS international research fellow (18F18315).
We thank a lot to Daniele Barducci,  Shinji Mukohyama, Marco Nardecchia for helpful conversations.
%. Y. -L. Zhang was supported by
%(B. Zhang was supported by ...).

\end{document}